\documentclass[fleqn,usenatbib]{mnras}

\usepackage{newtxtext,newtxmath}

\usepackage[T1]{fontenc}

\DeclareRobustCommand{\VAN}[3]{#2}
\let\VANthebibliography\thebibliography
\def\thebibliography{\DeclareRobustCommand{\VAN}[3]{##3}\VANthebibliography}

\usepackage{graphicx}	
\usepackage{amsmath}	

\title[Hydrostatic equilibrium of X-ray gas with HMG]{Hydrostatic equilibrium of X-ray gas in X-COP clusters with HMG}

\author[R. Monjo]{Robert Monjo\thanks{Email: \href{mailto:rmonjo@ucm.es}{rmonjo@ucm.es}}
\\
Department of Math and Computer Science, Saint Louis University,\\ Max Aub street, 5, E-28003, Madrid, Spain.
\\
Department of Algebra, Geometry and Topology, Complutense University of Madrid\\ Pza. Ciencias 3, E-28040 Madrid, Spain.
}

\date{Accepted XXX. Received YYY; in original form ZZZ}

\pubyear{\the\year{}}

\begin{document}
\label{firstpage}
\pagerange{\pageref{firstpage}--\pageref{lastpage}}
\maketitle

\begin{abstract}
Modified Newtonian Dynamics (MOND) was originally proposed to model galaxy rotation curves without dark matter. However, MOND presents difficulties in explaining the Radial Acceleration Relation (RAR) observed in galaxy clusters, and moreover, it does not completely eliminate the need for dark matter, since it requires using \textcolor{black}{non-luminous particles (e.g. cold molecular gas or durst, or neutrinos)} to explain the observed hydrostatic equilibrium of the hot gas. Hyperconical Modified Gravity (HMG) offers a relativistic framework that recovers the success of MOND in galaxy rotation curves as a natural particular case, and it could potentially reconcile the above discrepancies without invoking any type of dark matter. This Letter analyses the performance of the HMG model for hydrostatic equilibrium in \textcolor{black}{12 X-COP galaxy clusters, with a special focus on five objects with available complete stellar mass information:  A1795, A2029, A2142, A2319 and A644}. Specifically, we used high-resolution X-ray data with which gas density and mass profiles were previously derived to constrain modified gravity models. Our results show that the hydrostatic equilibrium of \textcolor{black}{analysed cluster gas is more naturally adjusted to the HMG model beyond 500 kpc without the need to fit parameters, but further research is required to expand to more spatial scales.}
\end{abstract}




\section{Introduction}
\label{sec:intro}

Classical models based on Newtonian gravity and general relativity (GR) suggest that visible matter in large gravitational systems (such as galaxy groups or clusters) is insufficient to account for observed gravitational effects, leading to the postulation of dark matter \citep{Zwicky1933,Clowe2006,vanDokkum2018,Maccio2020}. Despite the success of the dark matter paradigm in explaining a wide range of astrophysical phenomena, the nature of dark matter remains elusive \citep{Bertone2005,Akimov2011}. Direct detection experiments have yet to provide conclusive evidence of dark matter particles, and the discrepancies in cluster mass profiles warrant the exploration of alternative theories \citep{Okabe2010,Merritt2017,Roshan2021,Asencio2022,kroupa2023}. 

One of the supporting phenomena of the existence of dark matter is the hydrostatic equilibrium in galaxy groups and clusters, since it allows us to derive mass profiles from observable quantities such as gas density and temperature \citep{Angus2008,Biffi2016}. Under the assumption of hydrostatic equilibrium, the pressure gradient within the intracluster medium (ICM) balances the gravitational pull, leading to the following equilibrium condition: \begin{eqnarray} \label{eq:hydrost}
a_{in}(r) = \frac{1}{\rho(r)}\frac{dP(r)}{dr} = -\frac{GM(r)}{r^2} = a_{N}(r),
\end{eqnarray} where, $a_{in}(r)$ is the inward gravity, $a_N(r)$ is the Newtonian gravity, $P(r)$ is the pressure, $\rho(r)$ is the gas density, $G$ is the gravitational constant, $M(r)$ is the enclosed mass within the radial distance $r$ from the gravitational center.

This is a critical endeavor in astrophysics, as it provides key patterns about the fundamental forces and constituents of the universe. Galaxy clusters, being the largest gravitationally bound structures, offer a unique laboratory for testing theories of gravity and the nature of dark matter \citep{Humphrey2011, Humphrey2012, Horne2006}.  On the other hand, alternative theories such as the Modified Newtonian Dynamics (MOND), Moffat's Modified Gravity (MOG) and Hyperconical Modified Gravity (HMG) propose modifications to gravity that could potentially eliminate the need for dark matter \citep{Milgrom1983, Moffat2009, Milgrom2020, Monjo2023}. 

MOND was introduced by Milgrom in 1983 as an alternative to dark matter. MOND modifies Newton's second law for accelerations below a certain threshold \(a_0\), leading to the following effective gravitational acceleration: \begin{eqnarray}
\label{eq:mond}
a_{\text{MOND}} = a_{N}\,\nu(a_{N}/a_0) = \begin{cases}
a_{N} & \text{if } a_{N} \gg a_{0} \\
\sqrt{a_{N} a_0} & \text{if } a_{N} \ll a_{0}
\end{cases},
\end{eqnarray} where $a_0 \approx 1.2 \times 10^{-10} \, \text{m/s}^2$ is the characteristic acceleration scale, while $\nu(x)$ is an interpolation function that satisfies the transition between $\nu(x)=1$ for $x \gg 1$ and $\nu(x)=1/\sqrt{x}$ for $x << 1$ \citep{McGaugh2016}. \textcolor{black}{For example, the Mass-Luminosity-Scaling (MLS) function $\nu(x)= (1-\exp(-\sqrt{x}))^{-1}$ is empirically favoured according to a recent review \citep{Banik2022}}. MOND has been successful in explaining the flat rotation curves of spiral galaxies without invoking dark matter \citep{Sanders2003, McGaugh2007}. However, its application to galaxy clusters has been more challenging, often requiring additional unseen mass (from active or sterile neutrinos, or from cold baryonic particles) to fully account for the observed gravitational effects \citep{Angus2008, Banik2022}.

Hyperconical Modified Gravity (HMG) is a more recent proposal that modifies the gravitational potential by restricting GR to be only valid at a local scale over a background hypeconical metric \citep{Monjo2024}. HMG is a reativistic theory that reproduces MOND behavior with a unique natural transition function $\nu$ that addresses the shortcomings of MOND in explaining the radial acceleration relation (RAR) of galaxy clusters, \textcolor{black}{as \citep{MB2024} found using data of dynamical masses derived by \citet{Li2023}}. Specifically, the HMG transition function depends on the distribution of mass and the geometry of spacetime. 

To complete our previous findings, this \textit{letter} analyses the HMG prediction for hydrostatic equilibrium in \textcolor{black}{galaxy clusters with accurate X-ray gas measurements, obtained in XMM Cluster Outskirts Project (X-COP, \citealp[][]{Ghirardini2019}). The X-COP project measured the gravitational field of clusters up to large radii combining X-ray data from XMM-Newton with Sunyaev-Zeldovich (SZ) data from Planck \citep{Planck2014}. Thus, the hydrostatic mass profiles were extrapolated from the relationship between the gravitational field and the best-fitted X-ray temperature profiles, independently of other techniques such as gravitational lensing or cirular orbits. Therefore, the X-ray data provide us with highly accurate mass profiles of hot-gas baryonic matter. To reduce systematic errors, XMM-Newton's sensitivity to a broad range of X-ray energies makes it important to account for hydrogen column density ($N_H$) along the line of sight between the observer and the galaxy cluster, since high values of $N_H$ can cause significant attenuation in the soft X-ray regime ($0.1-2$ keV).} 

The hydrostatic mass profiles can differ from the \textit{caustic mass} estimated by gravitational lensing. In addition to the possible uncertainty caused by the parameter $N_H$, these differences could be mainly due to non-thermal pressure and the assumptions of hydrostatic equilibrium \citep{Andreon2017}. According to \citep{Maughan2016}, the mean bias between hydrostatic ($M_X$) and caustic ($M_C$) masses is $(M_X-M_C)/M_C = 0.20^{+0.13}_{-0.11}$. In any case, these observations have revealed strong discrepancies between the mass inferred from indirect techniques (such as the hydrostatic equilibrium or gravitational lensing) and the mass predicted by visible matter alone, suggesting the presence of dark matter or the need for modifications to our understanding of gravity. By comparing the predictions of gravity theories with these observational data, we can better understand their viability and constraints.

\section{Data and model}
\label{sec:obs_and_model}
\subsection{Observations used}
\label{sec:obs}
\color{black}

This \textit{letter} uses the published results of \citet{Eckert2022} on both the mass profiles and large-range inward gravity estimates of 12 X-COP clusters \citep{Ghirardini2019}, which are characterised by accurate X-ray gas measurements related to hydrostatic equilibrium in interstellar and intracluster mediums. Moreover, from these 12 galaxy clusters, five have available
complete stellar mass information: A1795, A2029, A2142, A2319 and A644 (see further details, for instance, in \citealp[][]{Eckert2022,Kelleher2024}).

In the context of the cold dark matter (CDM) paradigm, dark matter (DM) halos are expected to follow the so-called Navarro-Frenk-White (NFW) profile, which includes the characteristic parameters $R_{200}$ and $M_{200}$. The radius $R_{200}$ represents the radius within which the mean density of the cluster is 200 times the critical density $\rho_\text{crit}[z] = 3H^2[z]/(8\pi G)$ of the Universe at the cluster's redshift $z$ with Hubble parameter $H[z]$. This provides a standard definition for the extent of the cluster, encompassing the region where most of the cluster's mass is gravitationally bound. \citet{Eckert2022} used the estimation of the total mass within $R_{500}$, denoted $M_{500}$ to include both dark-matter and baryonic components from gas and stars (Table \ref{table:x_cop}).

\begin{table}
\centering
\caption{Properties of the X-COP galaxy clusters. The columns represent: the cluster name, redshift ($z$), hydrogen column density ($N_H$) in units of $10^{20}$ cm$^{-2}$, radius $R_{500}$ in kpc, and total mass $M_{500}$ in units of $10^{14} M_\odot$ (adapted from \citealp[][]{Eckert2022}). The [*] symbol indicates that available information on the different mass components is complete for these five galaxy clusters.}
\label{table:x_cop}
\begin{tabular}{lccccc}
\hline
Cluster & $z$ & $N_H$ & $R_{500}$ (kpc) & $M_{500}$ (10$^{14} M_\odot$) \\
\hline
A85    & 0.0555 & 2.8  & $1214^{+4}_{-4}$   & $5.03^{+0.05}_{-0.05}$ \\
A644\,[*]    & 0.0704 & 7.5  & $1398^{+16}_{-16}$ & $7.80^{+0.26}_{-0.27}$  \\
A1644  & 0.0473 & 4.1  & $1031^{+11}_{-10}$ & $3.06^{+0.09}_{-0.09}$  \\
A1795\,[*]  & 0.0622 & 1.2  & $1160^{+6}_{-6}$   & $4.42^{+0.07}_{-0.07}$   \\
A2029\,[*]   & 0.0766 & 3.2  & $1340^{+9}_{-9}$   & $6.91^{+0.14}_{-0.14}$  \\
A2142\,[*]   & 0.09   & 3.8  & $1453^{+9}_{-9}$   & $8.93^{+0.17}_{-0.16}$  \\
A2255  & 0.0809 & 2.5  & $1202^{+17}_{-16}$ & $5.01^{+0.22}_{-0.20}$ \\
A2319\,[*]   & 0.0557 & 3.2  & $1424^{+5}_{-4}$   & $8.12^{+0.08}_{-0.08}$  \\
A3158  & 0.059  & 1.4  & $1146^{+12}_{-11}$ & $4.25^{+0.14}_{-0.13}$  \\
A3266  & 0.0589 & 1.6  & $1381^{+10}_{-10}$ & $7.43^{+0.16}_{-0.15}$ \\
RXC1825 & 0.065  & 9.4  & $1109^{+9}_{-8}$   & $3.88^{+0.09}_{-0.08}$  \\
Zw1215  & 0.0766 & 1.7  & $1346^{+21}_{-21}$ & $7.00^{+0.35}_{-0.32}$   \\
\hline
\end{tabular}
\end{table}

\color{black}

\subsection{Aceleration from hyperconical modified gravity (HMG)}

Inward gravity observed in the four considered systems was compared to the predicted acceleration of HMG, given by visible matter, according to the work developed by \citet{Monjo2023, MB2024} and summarised here. Let $g$ be the background metric of the hyperconical universe \citep{Monjo2017,MCS2023}. The metric $g$ is locally approximately given by \begin{equation} 
g \approx dt^2 \left(1- kr'^2 \right) 
-  \frac{t^2}{t_{0}^2} \left( \frac{dr'^2}{1-kr'^2} + {r'}^2d{\Sigma}^2 \right)
- \;\; \frac{ 2r't}{t_{0}^2} \frac{dr'dt}{\sqrt{1-kr'^2}}\,,
{} \label{eq:inhom0}
\end{equation} where $k = 1/t_0^2$ is the spatial curvature for the current value of the age $t$ of the universe ( $t_0 \equiv 1$), while $t/{t_0}$ is a linear scale factor, $r' \ll t_0$ is the comoving distance, and $\Sigma$ represents the angular coordinates. The shift and lapse terms of Eq. \ref{eq:inhom0} lead to an apparent radial spatial inhomogeneity that is assimilated as a fictitious acceleration with adequate projected coordinates.

\color{black}

Then, the HMG model is based on three key ideas:
\begin{itemize}
    \item[i)] any gravitational system of mass $M_{sys}(r)$ generates a perturbation over the background metric $g \to \hat{g}$ in Eq. \ref{eq:inhom0} such that \begin{eqnarray}
    k{r'^2} \to \hat{k}{{\hat r}'^2} := k{{\hat r}'^2} +2\int_{{\hat r}'}^\infty \frac{GM_{sys}(r)}{c^2 r^2} dr\,
    \end{eqnarray}
   \item[ii)] GR is only (locally) valid with respect to the background metric, that is, gravity is due to the perturbation term $\hat{h} := \hat{g}-g$;
   \item[iii)]  the coordinates that parameterise the metric, with apparent radial inhomogeneity, are projected as follows:\begin{eqnarray} \nonumber
    r' \;\; &\to& \;\; {\hat r}' = \lambda^{1/2}r' 
    \\ \nonumber
    t \;\; &\to& \;\; \hat t = \lambda t
\end{eqnarray} where $\lambda := 1/(1-\gamma/\gamma_0)$ is the scaling factor, which is a function of the angular position $\gamma = \sin^{-1}(r'/t_0)$ and a projection factor $\gamma_0^{-1} = \gamma_{sys}^{-1}\cos \gamma_{sys}$, where $\gamma_{sys}$ is the characteristic angle of the gravitational system. In an empty universe, $\gamma_0 = \gamma_U / \cos \gamma_U$. We expect $\gamma_U = \frac{1}{3}\mathrm{\pi}$ and therefore that $\gamma_0 = \frac{2}{3}\mathrm{\pi} \approx 2$ \citep{MCS2023}.
\end{itemize}

\color{black}

When geodesic equations are applied to the projected time component of the perturbation $\hat{h}_{tt}$, a fictitious cosmic acceleration of roughly $\gamma_0(r)^{-1}c/t = \gamma_{sys}^{-1}(r)\cos\gamma_{sys}(r)\; c/t$ emerges in the spatial radial direction. Specifically, the \textbf{spatial contribution} to the acceleration derived from $\hat{h}_{tt}$ is \citep{MB2024}:\begin{eqnarray} 
\label{eq:RAR_aN}
      a_{r}(r) \approx a_N(r) + \frac{c}{\gamma_0(r) t}\,,
\end{eqnarray}where $a_N(r) := {G}M_{sys}/r^2$ is the Newtonian acceleration. However, time-like component is not negligible and it contributes to the total centrifugal acceleration $a_{C}(r)$ such that \citep{Monjo2023}:\begin{eqnarray}
\label{eq:centrifugal}  \nonumber
    {} & a_{C}(r)^2 \approx & a_N(r)^2 + |a_N(r)|\frac{2c}{\gamma_0(r) t} \;\;\Longrightarrow\;\;
    \\
   & \Longrightarrow & a_{C}(r) \approx a_N(r)\sqrt{1 + \frac{a_{\gamma 0}(r)}{|a_N(r)|}}\,,
\end{eqnarray}with background cosmic acceleration $a_{\gamma 0}(r) := 2\gamma_0(r)^{-1}c/t$ that contributes to a transition function $\nu(a_N,a_{\gamma 0}) := \sqrt{1+a_N/a_{\gamma 0}}$ depending on $a_N$ like the Milgromian gravity but also on the radial position $r$ from the projection factor $\gamma_0^{-1}(r) = \gamma_{sys}(r)^{-1}\cos \gamma_{sys}(r)$ that is key in $a_{\gamma 0}(r)$. The projective angle $\gamma_{sys}(r)$ can be estimated from a general approach by considering the relative geometry (angle) between the Hubble speed $v_H(r) := r/t$ and the Newtonian circular speed $v_{N}(r) := \sqrt{{G}M_{sys}(r)/r}$, as follows \citep{MB2024}: \begin{eqnarray}
\label{eq:model_general2} \nonumber
{} & \sin^2 \gamma_{sys}(r) \;\; \approx  \;\; 
\\
 \approx  & \sin^2\gamma_U + \left({\sin^2\gamma_{center} -  \sin^2 \gamma_U} \right) \bigg| \frac{2v_N^2(r)-\epsilon^2_H v_H^2(r)}{2v_N^2(r) + \epsilon^2_H v_H^2(r)}\bigg|\;,
\end{eqnarray}where the parameter 
$\epsilon^2_H$ is the so-called \textit{relative density of neighborhood}, while $\gamma_{center} \approx \mathrm{\pi}/2$ and $\gamma_U = \mathrm{\pi}/3$ can be fixed here to set a 1-parameter ($\epsilon_H$) general model from Eq. \ref{eq:model_general2}. However, the parameter $\epsilon_H$ is not totally free, since it is theoretically and statistically strongly linked to the observed density $\rho(r)$ of the cluster at $r \sim 50-500\,$kpc, which corresponds to a reference of dynamical equilibrium \citep{MB2024}. 
\color{black}
In particular, if we assume such equilibrium for our considered distances, it is $\epsilon_{H}^2(r) \approx {\rho(r)}/{\rho_{vac}} + \frac{1}{6} \approx {\rho(r)}/{\rho_{vac}}$, where $\rho_{vac} := 3/(8\pi G t^2)$ is the HMG vacuum density\textcolor{black}{, which coincides with the critical density $\rho_\text{crit}$ for a linear expanding Hubble parameter $H = 1/t$}. In this case, $\gamma_{sys}(r) = \gamma_U \equiv \frac{1}{3}\pi$ and $\gamma_0 = \gamma_U /\cos \gamma_U = 2\pi/3$ are constant. In general terms, we will assume that the equilibrium is reached only at $r \sim 50-500\,$kpc and, therefore, we use the following: \begin{eqnarray}
\label{eq:epsil}
    \epsilon_{H}^2(r) \approx \begin{cases} 
    {\rho(50\text{kpc})}/{\rho_{vac}} & \text{if}\;\ r < 50 \text{kpc}\\
    {\rho(r)}/{\rho_{vac}} & \text{if}\;\ 50 \text{kpc} < r < 500 \text{kpc}\\
    {\rho(500\text{kpc})}/{\rho_{vac}} & \text{if}\;\; r > 500 \text{kpc}
\end{cases}  \,.
\end{eqnarray}
Notice that the range defined by $ 50\,$kpc and $500\,$kpc could be adjusted for each cluster but, to reduce over-fitting in the results, we preferred to keep the values found in previous studies.

Therefore, Eqs. \ref{eq:RAR_aN} and \ref{eq:centrifugal} are considered dependent on the projection factor $\gamma_0^{-1}(r) = \gamma_{sys}(r)^{-1}\cos \gamma_{sys}(r)$ according to Eqs. \ref{eq:model_general2} and \ref{eq:epsil}.

Finally, the equations of hydrostatic equilibrium for HMG are obtained from the classical hydrostatic equilibrium (Eq. \ref{eq:hydrost}) but replacing the Newtonian gravitational acceleration by the total HMG centrifugal acceleration (Eq. \ref{eq:centrifugal}), that is, \begin{eqnarray}
a_{in}(r) = \frac{1}{\rho(r)}\frac{dP(r)}{dr} \approx a_N(r)\sqrt{1 + \frac{a_{\gamma 0}(r)}{|a_N(r)|}},\end{eqnarray} where the inward acceleration $a_{in}(r)$ is estimated from observed X-ray temperature profiles (Sec. \ref{sec:obs}).
\color{black}

\section{Results and discussion}
\label{sec:results}

\begin{figure*}
\centering
	\includegraphics[width=0.99\textwidth]{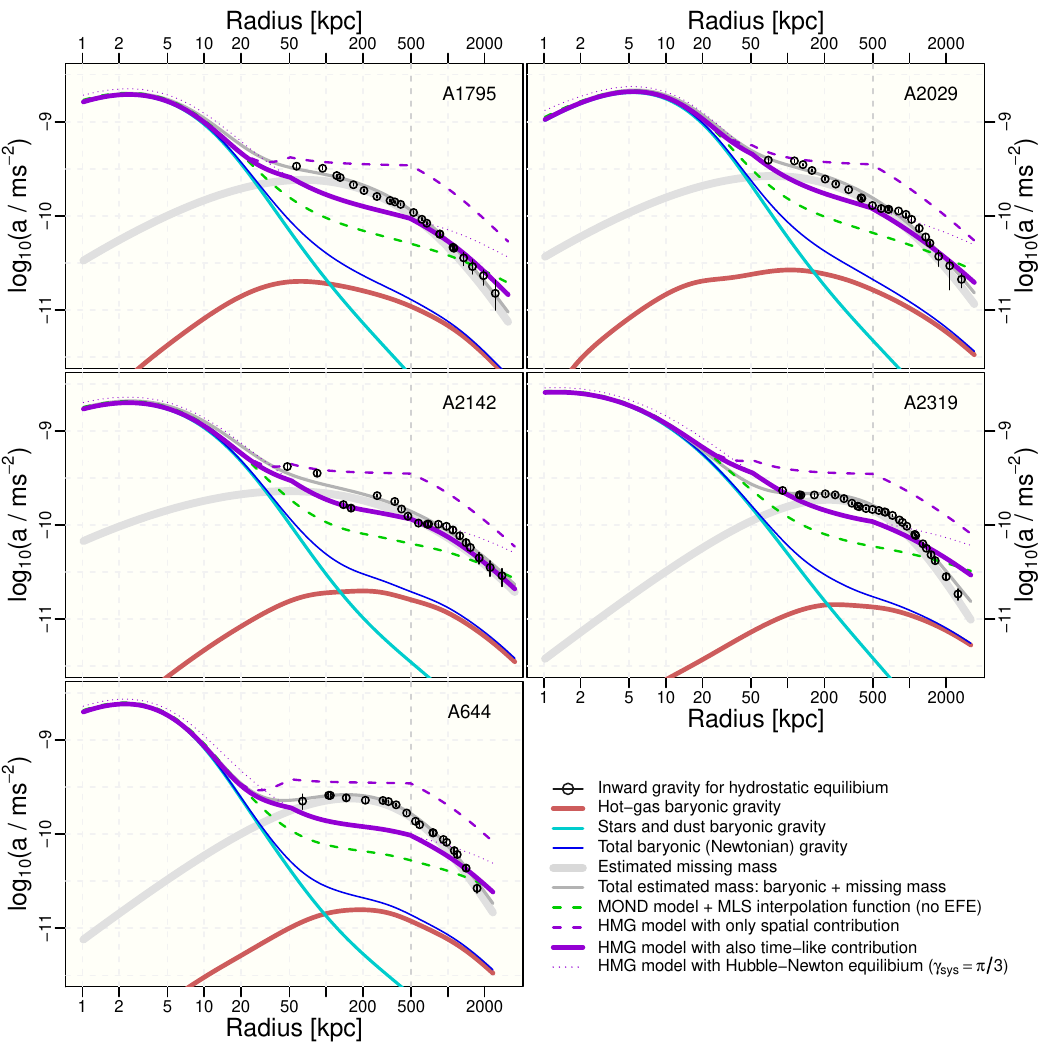}
    \caption{Acceleration profiles for the five X-COP galaxy clusters that have complete information on mass profiles: star baryonic matter (light blue line), hot-gas baryonic matter (red line) and total baryonic matter (dark blue kine). The \textit{missing mass} (thick gray line) was estimated by \citealp[][]{Eckert2022} from the hot-gas X-ray spectrum, accounting for the inward gravity required for hydrostatic equilibrium (circles). MOND prediction with the MLS interpolation function (dashed green line) is compared to the HMG prediction for the purely radial spatial component (Eq. \ref{eq:RAR_aN}, dashed purple line) and total acceleration (Eq. \ref{eq:centrifugal}, thick purple line). The balanced case between Hubble flux and Newtonian orbital speed is represented by the dotted purple line. A vertical line is shown at 500 kpc to highlight the range from which hot-gas gravity is dominant.
}  \label{fig:Fig01}
\end{figure*}

\begin{figure*}
\centering
	\includegraphics[width=0.99\textwidth]{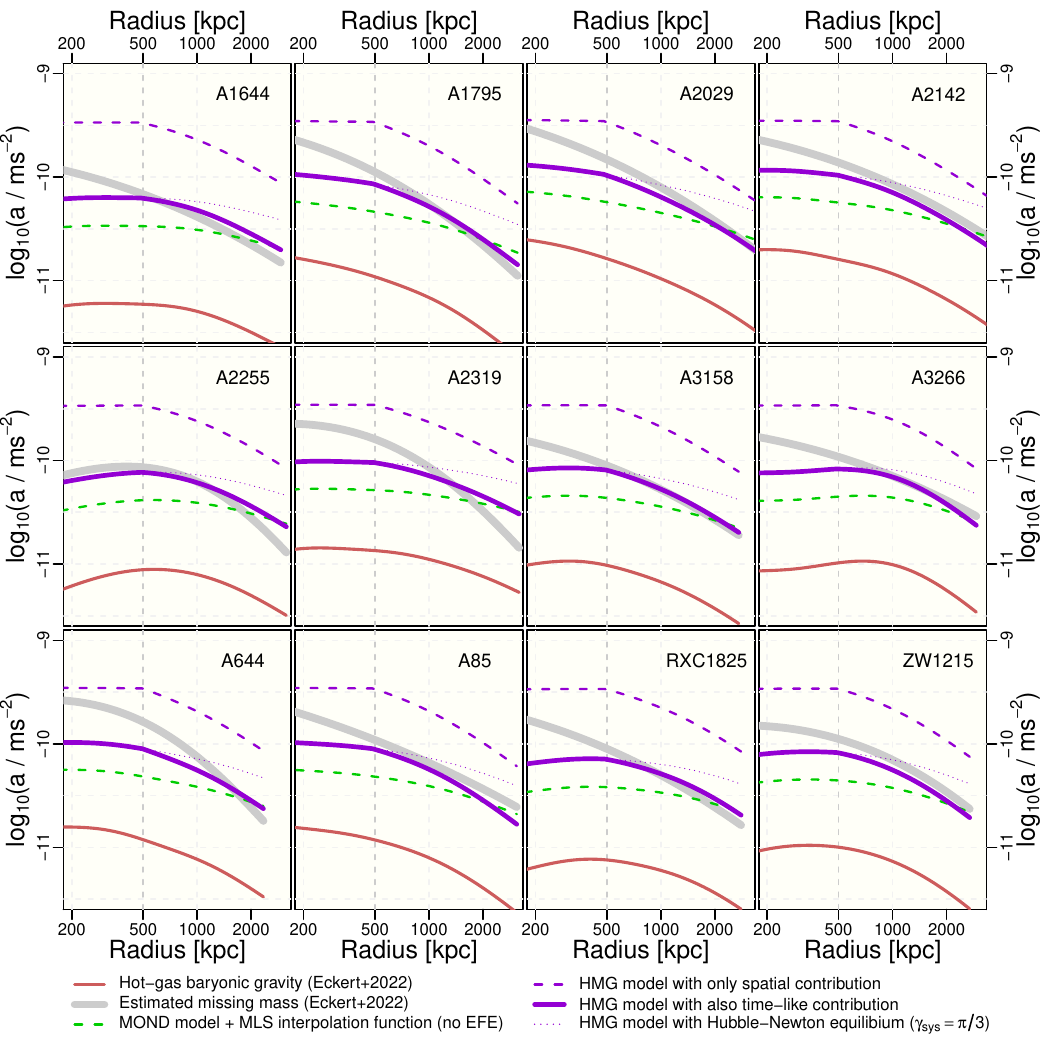}
    \caption{Acceleration profiles for the 12 X-COP galaxy clusters that have accurate data of hot-gas baryonic matter (red line) and  \textit{missing mass} (thick gray line) estimated by \citealp[][]{Eckert2022} taking into account the inward gravity required for hydrostatic equilibrium. MOND prediction with the MLS interpolation function (dashed green line) is compared to the HMG prediction for the purely radial spatial component (Eq. \ref{eq:RAR_aN}, dashed purple line) and total acceleration (Eq. \ref{eq:centrifugal}, thick purple line). The balanced case between Hubble flux and Newtonian orbital speed is represented by the dotted purple line. As in \ref{fig:Fig01}, a vertical line is shown at 500 kpc to highlight the range from which hot-gas gravity is dominant.
}  \label{fig:Fig02}
\end{figure*}

\textcolor{black}{The results showed an accurate concordance between the acceleration predicted by the HMG model and the inferred from hot-gas mass profiles published by \citet{Eckert2022}. Specifically, by assuming the simplest HMG version that describes dynamical equilibrium ($\gamma_{sys} = \gamma_U = \pi/3$), the predicted acceleration is closer to the estimates of inward gravity necessary for the hot-gas hydrostatic equilibrium (Fig. \ref{fig:Fig01} and \ref{fig:Fig02}), reducing the need for or without requiring additional matter content (e.g., exotic dark matter, cold baryonic particles, or active/sterile neutrinos), in contrast to MOND theories \citep{Angus2008, Banik2022, Kelleher2024}. In particular, the validity of the spatial ranges analysed extends beyond 500 kpc, reaching 2Mpc according to the X-ray emission from the hot intracluster medium \citep{Stott2008, Eckert2022, Kelleher2024}. However, a discrepancy between total inward gravity and the HMG prediction is found in the 50-500 kpc range (Fig. \ref{fig:Fig01}), probably due to some uncounted baryonic mass (e.g., cold molecular clouds) or a more complex casuistic misunderstood by the HMG model.}


\textcolor{black}{Using X-COP data as in the present work, \citet{Kelleher2024} also needed to consider a fitted additional (hypothetical) component for similar spatial ranges. They described that this component may consist of small and compact clouds of atomic hydrogen (with poor X-ray emission). Moreover, \citet{Kelleher2024} obtained good results by assuming an External Field Effect (EFE) in the gravitational strength within the MOND paradigm. It was also inferred by fitting to the X-ray data, but they support the outcome to be consistent with independent estimates from the large-scale structure of the Universe.}

The good performance of the HMG model beyond 500 kpc is actually remarkable in comparison to earlier work in MOND. For instance, \citet{Angus2008} tested the compatibility of the required dark matter with ordinary massive neutrinos at the experimental limit of detection at that moment ($m_\nu = 2$ eV) and showed that they cannot account for the required amount within regions with temperatures lower than 2 keV. This study concluded that cluster baryonic dark matter or sterile neutrinos would be a more satisfactory solution to the problem of missing mass in MOND X-ray emitting systems. The current upper bound for the neutrino mass is $m_\nu < 0.8$ eV according to \citet{katrin2022} and the sum of masses of $\sum m_\nu < 0.12$ eV was inferred from \citet{Planck2020b}.

Other modified gravity theories were also explored to solve this problem. For example, the study of \citet{Horne2006} provides significant insight into inward acceleration anomalies by applying conformal gravity models to the Abell 2029 cluster \citet{Horne2006}. Horne examined the hydrostatic equilibrium of the X-ray gas in Abell 2029 using high-resolution Chandra data, probing the gravitational potential on scales much larger than those accessible with traditional methods. His analysis revealed that conformal gravity could also bind the X-ray gas without requiring dark matter, although this model predicted too much gravity in certain regions.

\section{Concluding remarks}
\label{sec:conclusions}

In this study, we have explored the viability of HMG in explaining the hydrostatic equilibrium of X-ray gas in \textcolor{black}{X-COP galaxy clusters with a special focus on five cases with complete mass profiles}. Our analysis, based on high-resolution X-ray data, demonstrates that HMG successfully accounts for the observed mass distribution in hot gas \textcolor{black}{beyond 500 kpc without invoking additional missing mass (e.g. ordinary non-luminous matter)}, which is typically required in MOND theories. \textcolor{black}{However, a significant discrepancy is still produced between the HMG prediction and observation for the 50-500 kpc range.}

\textcolor{black}{The uncertainty levels on the observed inward gravity and baryonic components are small compared to that discrepancy and to the components themselves. Therefore, in agreement with the findings of \citet{Kelleher2024}, a possible systematic uncertainty in the modelling of modified gravity (e.g. MOND or HMG) would be related to its strong dependence on possible unknown baryonic components in addition to the matter already considered (i.e., stars, durst, and gas), such as small but compact clouds of atomic hydrogen. }

This suggests that HMG offers a promising alternative framework for understanding the gravitational dynamics in galaxy clusters. Further research and observations are necessary to refine these models and extend their application to a broader range of astrophysical systems. We suggest evaluating the effectiveness of HMG in explaining the mass distribution in galaxy clusters and the dynamics or growth of larger cosmic structures.

\section*{Acknowledgements}

We greatly appreciate the very useful data published by \citet{Eckert2022}. We would like to highlight the collaborative efforts of colleagues in the astrophysics community with a special acknowledgment to F. Lelli and D. Eckert, whose insights and feedback have been invaluable.

\section*{Data Availability}

In this study, no new data was created or measured.



\bibliographystyle{mnras}
\input{00_mnras.bbl}








\bsp	
\label{lastpage}
\end{document}